\documentclass[12pt]{article}
\usepackage{graphicx}
\usepackage{amsmath}
\usepackage{amssymb}
\usepackage{psfrag}
\usepackage{pslatex}
\usepackage{color}
\usepackage{times}

\newlength{\figWidth}
\newcommand{\dt}{\delta t}
\newcommand{\dtr}{\delta t_r}
\newcommand{\dts}{\delta t_\sigma}

\newcommand{\price}{\text{price}}
\newcommand{\yearUnit}{\text{year}}
\newcommand{\gr}{\text{gr}}
\newcommand{\Prob}{\text{Pr}}
\newcommand{\E}[1]{ E\left[~ #1 ~\right]}
\newcommand{\avg}[1]{\langle #1 \rangle}

\begin{document}
\begin{center}

\thispagestyle{empty}
\setcounter{page}{0}
{\Huge\bf Time reversal invariance in finance}\\[2ex]


\vspace{2ex}
{\large\bf Gilles Zumbach}\\[4ex]
\parbox{0.8\textwidth}{RiskMetrics, Av des Morgines 12, 1213 Petit-Lancy, Switzerland.\\[1ex]
e-mail: gilles.zumbach@riskMetrics.com
}\\[2ex]
and\\[2ex]
\parbox{0.8\textwidth}{Consulting in Financial Research,
Ch. Charles Baudouin 8,\\
1228 Saconnex d'Arve, Switzerland.\\[1ex]
e-mail: gilles.zumbach@bluewin.ch

}

\vspace{5ex}
January 2007

\vspace{6ex}
{\Large\bf Abstract}
\end{center}

Time reversal invariance can be summarised as follows: 
no difference can be measured if a sequence of events is run forward or backward in time. 
Because price time series are dominated by a randomness that hides possible structures and orders,
the existence of time reversal invariance requires care to be investigated. 
Different statistics are constructed with the property to be zero for time series 
which are time reversal invariant;
they all show that high-frequency empirical foreign exchange prices are not invariant.
The same statistics are applied to mathematical processes that 
should mimic empirical prices. 
Monte Carlo simulations show that only some ARCH processes with a 
multi-timescales structure can reproduce the empirical findings.
A GARCH(1,1) process can only reproduce some asymmetry.
On the other hand, all the stochastic volatility type processes are time reversal invariant.
This clear difference related to the process structures gives some strong selection criterion
for processes.

\vspace{2ex}

\newpage
\section{Introduction}
Time reversal invariance (TRI) is a very important concept in science. 
The idea can be summarised as follows: when a sequence of events is viewed starting from the end, 
namely with the arrow of time reversed, 
is it possible to measure a difference compared to the normal time ordering?
A rigorous formulation of time reversal invariance is that the transformation $t \rightarrow -t$ is an exact symmetry of the system under consideration. 
The basic laws of physics are time reversal invariant 
(Newton equation for mechanic, Maxwell equations for electromagnetism, Einstein equation for general relativity, Dirac equation for quantum mechanics, etc...),
but the macroscopic world is clearly not time reversal invariant. 
This paradox was solved by thermodynamics and the increase of entropy.
The same question can be asked about finance, namely if a time series of prices originating in the financial market is time reversed, can we ``see'' the difference? 
Contrarily to a Buster Keaton movie, at the level of prices or returns, 
it is indeed very difficult to notice a difference
because financial time series are dominated by a randomness 
that hides possible structures and orders. 
Therefore, the appropriate formulation of the question is whether statistics 
can show the presence, or absence, of time reversal invariance. 

\cite{RamseyRothman.1988, RamseyRothman.1996} have already addressed 
this question in an economic framework.
Their idea is to search for differences between up and down moves over long time horizons, 
using yearly economic indicators. 
This behaviour is typically related to business cycles, 
for example with long slow rises followed by abrupt decreases.
The estimator that is used is given by $\E{r^2(t)~r(t+k\dt)}$ 
where $\dt$ is the time increment of the time series, 
$r$ the return and $k \neq 0$ an integer index. 
As the amount of long term economic data is fairly small, 
such studies need to rely on a carefull analysis of the statistical properties of the indicators
\cite{RamseyRothman.1996}.
Other indicators have been proposed \cite{Chen.2000, Fong.2003}.
The salient results is that most of the economic time series are not time reversal invariant,
but for a small fraction of them, the null hypothesis of time reversal invariance cannot be rejected.

This paper takes a different angle to study TRI in empirical data by using 
high-frequency foreign exchange time series.
The focus is to study various statistics related to the volatility, 
for time horizons ranging from 3 minutes to 3 months.
For foreign exchange rates, a symmetry between the exchanged and expressed currencies is plausible
(at least for major free floating currencies).
This symmetry occurs because a FX rate is a conversion factor between two numeraires, 
and not the price of a security expressed in a numeraire (like for an equity price or of a bond price).
Under the exchange of the currencies, an exchange rate is transformed by $\price \rightarrow 1/\price$,
and the logarithmic returns by $r \rightarrow -r$.
Notice that reversing the time induces the same transformation  $r \rightarrow -r$
and the reverse ordering of the time series.
Therefore, if the exchange of currencies is an exact symmetry, then all statistics 
that are odd\footnote{An even function is such that $f(-r) = f(r)$, an odd function such that $f(-r) = - f(r)$}
in the returns are zero, like for example $\E{r^2(t)~r(t+k\dt)}$.
The same argument implies that the return probability distribution in even, that is,  $p(r) = p(-r)$.
We have checked below the empirical validity of this argument for $p(r)$, 
and found that it is likely incorrect.
Even if ultimately incorrect, the argument points to a small asymmetry for such term. 
In order to have a better signature of time irreversibility, 
a better track for foreign exchange data is to search for 
estimators that are even in the returns, 
but sensitive to time reversal.

A general definition of TRI is given for example in \cite{Chen.2000}. 
In particular, they show that the distribution of the returns must 
be symmetric for a TRI series, 
and they construct a test based on this property using the characteristic function of the pdf.
A more general construction along this line is as follow.
Take a quantity $\sigma$ 
such that $\sigma \rightarrow \sigma$ under the time reversal transformation.
For example $\sigma$ can be the logarithmic price;
in this paper $\sigma$ is a volatility estimator computed from the returns.
Then, the quantity $\Delta\sigma(t) = \sigma(t+\dt) - \sigma(t)$ is 
odd under the time reversal transformation: $\Delta\sigma \rightarrow -\Delta\sigma$.
If the series is TRI, then the distribution of $\Delta\sigma$ must be even.
Another idea for testing TRI is based on covariance or 
correlation between two quantities $x$ and $y$.
Essentially, the quantity $\E{x(t)~y(t+\dt)}$ should 
be equal to $\E{x(t+\dt)~y(t)}$ if the series is TRI,
and a test can be constructed on the difference between these two quantities.
Because of the symmetry between the arguments of the covariance or the correlation, 
the two quantities must be different in order to have a non trivial test.
\cite{RamseyRothman.1988} use $r$ and $r^2$;
this paper uses volatilities $\sigma$ with different parameters.

This paper presents three different statistical estimators sensitive 
to TRI.
These statistics are based on various estimator of the volatility (i.e. even in the returns), 
and are essentially measuring time reversal for volatilities computed
with information before $t$ and after $t$. 
They all show that empirical data are clearly not time reversal invariant.

Using the same statistical tool, 
various processes that should mimic empirical data are investigated. 
For example, a simple Gaussian random walk is time reversal invariant, 
and the three statistics are zero. 
Much more interesting are the processes from the ARCH family and from the stochastic volatility family.
Some of the ARCH type processes can reproduce the empirical figures,
but all the simulated stochastic volatility processes are time reversal invariant.
These results demonstrate that the last family of processes cannot describe 
some stylized facts of financial time series.
It is an important results as TRI statistics allow us to select 
between models that are structurally very different.
The usual method for model selection is to nest the processes and to show that the corresponding 
parameter is significantly different from zero.
But this method is not generally possible for processes 
that are so widely different as ARCH and stochastic volatility.  
Besides, the systematic comparison of the results between empirical data and processes 
allow us to glimpse the origin of time asymmetry.
This discussion is presented just before the conclusion.  

\section{Data and notations}
The empirical data used in the empirical study originate from the foreign exchange market.
The high-frequency tick-by-tick quotes are used to compute a continuous ``business time scale'' 
\cite{Breymann.2000} in order to deseasonalize the strong daily and weekly patterns present in such data.
The prices are sampled each 3 minutes in business time in order to obtain 
deseasonalized homogeneous time series.
An efficient deaseasonalisation procedure is crucial for the presented computations,
as the use of a 3 minutes regular sampling in physical time (after the week-ends have been removed)
would mainly show the strong daily pattern related to the opening and closing of the various market
around the world. Such strong seasonality would hide other interesting stylized facts.
The author is gratefull to Olsen \& Associates in Zurich for providing the deseasonalized data.
The in-sample used in this study starts January 1, 1990 and ends July 1, 2001.
The year 1989 is used to build-up the computations.

Our notations are as follows: 
$r$ denotes a time series and $r(t)$ the value of the time series at time $t$.
The parameters are denoted between bracket, like for the time series $r[\dtr]$ 
or for the value $r[\dtr](t)$ at time $t$.
The subscript are used to denote different estimators, for example $\sigma_h$ and $\sigma_r$.
The raw data are given by the logarithmic price time series $x(t)$ defined between a 
start time $t_s$ and an end time $t_e$.
The time reversal transformation corresponds to changing $x(t_s + \Delta T)$ by $x(t_e - \Delta T)$,
which is denoted informaly as $x(t) \rightarrow x(-t)$.. 
The return $r$ over a time interval $\dtr$ 
is defined by $r[\dtr](t) = x(t) - x(t-\dtr)$.

The historical volatility $\sigma_h$ is
\[
    \sigma_h^2[\dts, \dtr](t) = 
	\frac{1 \yearUnit}{\dtr} ~\frac{1}{n} ~\sum_{t-\dts+\dtr \leq t' \leq t} r^2[\dtr](t').
\]
Essentially, $\sigma_h(t)$ measures the fluctuation of the 
prices at the scale $\dtr$, in the time interval $[t-\dts, t]$.
This definition includes information only in the past of $t$. 
The ratio $1 \yearUnit/\dtr$ annualises the volatility, and $n$ is the number of terms in the sum over $t'$.
In the actual evaluation, the sum over $t'$ is carried over each point on the 3 minutes time grid.

The realized volatility $\sigma_r$ is essentially the same definition, 
but using information only in the future of $t$
\[
    \sigma_r[\dts, \dtr](t) = 
	\frac{1 \yearUnit}{\dtr} ~\frac{1}{n} ~\sum_{t+\dtr \leq t' \leq t+\dts} r^2[\dtr](t') 
	= \sigma_h[\dts, \dtr](t+\dts).
\] 
Notice that a definition of volatility depends on two time intervals, 
namely the time interval $\dtr$ over which the returns are computed (also called the granularity) 
and the time interval $\dts$ over which the returns variance is computed.  
In order to have a good volatility estimator, 
the ratio $\dts/\dtr$ should be large enough, say $\dts/\dtr > 10$.
Except for the third time asymmetry estimator, this ratio is fixed to $\dts/\dtr = 24$.

\section{Empirical time reversal statistics}
\begin{figure}
  \centering
  \psfrag{x}{$\Delta\sigma$}
  \psfrag{y}{$p(\Delta\sigma)$}
  \includegraphics[width=\figWidth]{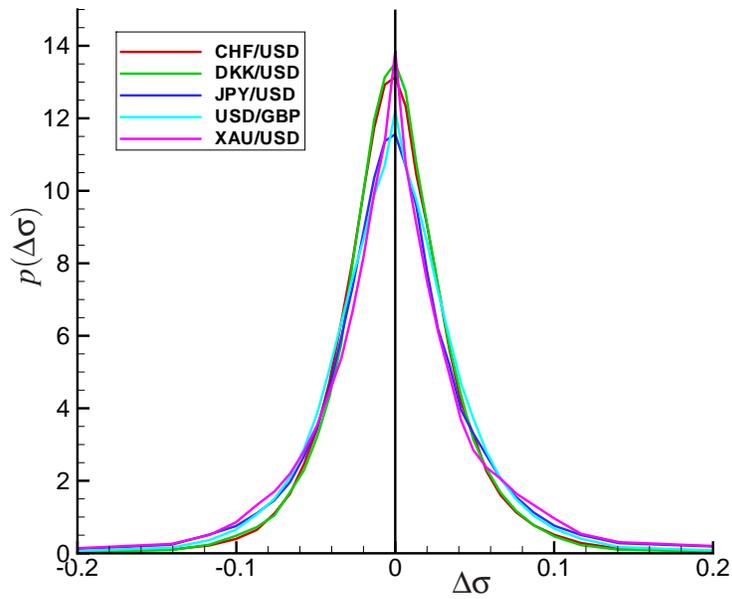}
  \caption{\it Probability density for the volatility increment $\Delta\sigma$, 
   for the foreign exchanges CHF/USD, DKK/USD, JPY/USD, USD/GBP and for gold XAU/USD. 
   The time horizons are $\dts = $1 day, $\dtr = \dts/24$. 
   Similar figures are obtained for other time horizons.}
  \label{fig:DV_PDF}
\end{figure}

The changes of the volatility are measured by the volatility increment 
\[
   \Delta \sigma = \sigma_r - \sigma_h. 
\]
With a time reversal transformation, 
the volatility increment changes by $\Delta \sigma(t) \rightarrow -\Delta \sigma(t)$. 
The probability density $p(\Delta\sigma)$ of the volatility increment can be estimated
and its asymmetry $a_p(\Delta \sigma) = p(\Delta \sigma) - p(-\Delta \sigma)$ 
gives a measure of the time irreversibility. 
This asymmetry is a quantitative measure of the following intuitive perception of the price dynamics. 
A shock on the market (for example due to the arrival of an important piece of news) produces a sudden increase of volatility, followed by a slow relaxation toward a normal level of volatility. 
For the time reversed series, this corresponds to a slow volatility increase followed by a sudden return to the normal and the distribution of volatility increments is $p(-\Delta \sigma)$. 
The asymmetry $a_p(\Delta \sigma)$ measures the asymmetry in the dynamics between the original and time reversed series. 
Figure~\ref{fig:DV_PDF} shows the probability distribution $p(\Delta \sigma)$.
The probability is estimated by binning the empirical data using a 
linear interpolation in a non uniform grid.
The points on the sampling grid are chosen so 
that a roughly equal number of values fall into each bins.   

\begin{figure}
  \centering
  \psfrag{v}{$\Delta\sigma$}
  \psfrag{w}{$a_p(\Delta\sigma)$}
  \includegraphics[width=\figWidth]{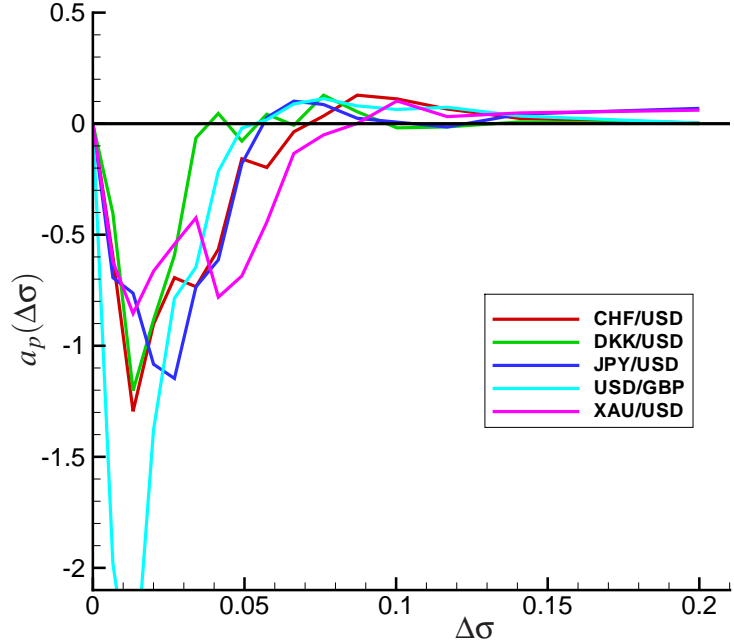}
  \caption{\it Asymmetry of the probability density for the volatility increment $\Delta\sigma$.
   The parameters and time series are as for fig.~\ref{fig:DV_PDF} }
  \label{fig:DV_PDF_asymmetry}
\end{figure}

The pdf appears to be symmetric at first glance.
It shows that our intuitive perception as described above is exaggerated. 
Yet, a detailed examination of $p(\Delta \sigma)$ reveals the expected asymmetry
around $\Delta \sigma \simeq 0.02$. 
Figure~\ref{fig:DV_PDF_asymmetry} displays the asymmetry $a_p(\Delta \sigma)$, 
and a fairly consistent symmetry breaking pattern for various empirical time series is observed.
The negative values for $\Delta\sigma \lesssim 0.05$  
corresponds to the ``return to the normal'' 
or to a larger probability for small negative volatility increments.
There is a corresponding larger probability for large volatility increments $\Delta \sigma$ 
(that is the arrival of news or shocks), 
that translate into positive values for $a_p$.
Because the volatility is stationary, 
a simple empirical first moment like $\avg{\Delta \sigma}$ converges toward zero 
with the inverse of the sample size.
This shows that more complex statistics should be used to reveal this asymmetry.

\begin{figure}
  \centering
  \includegraphics[width=0.80\textwidth]{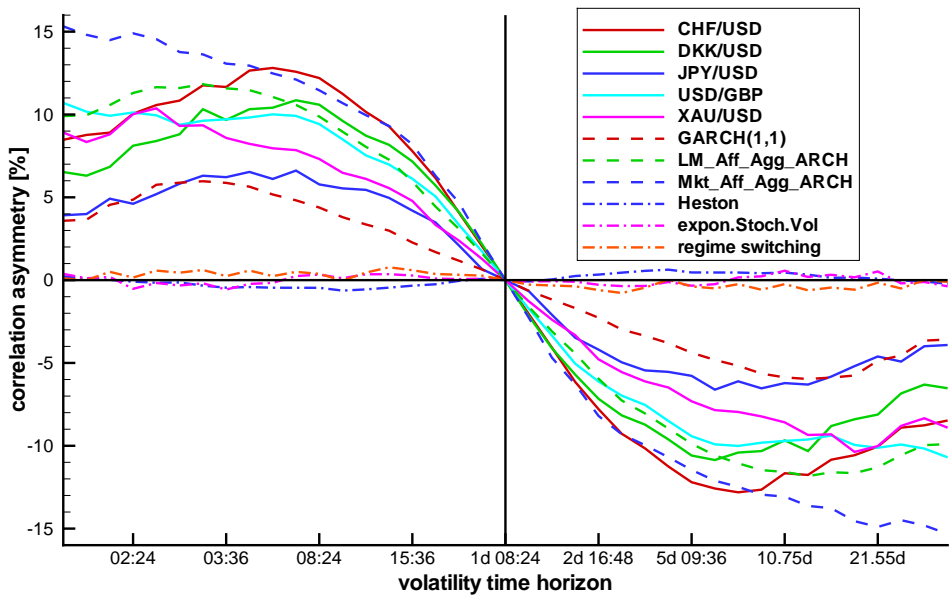}
  \caption{\it The measure of asymmetry $a_\sigma(\dts, \dts')$.
   The parameter $\dts'$ is given on the horizontal axis, 
   the value of $\dts$ is the symmetric through the vertical axis 
   (for example for $\dts'$ = 8h24 corresponds $\dts$ = 5 day 9h36).
   }
  \label{fig:antisymCut_volH_volR}
\end{figure}
The second statistics involves the correlation of volatilities at various time horizons. 
The following correlation is investigated in \cite{ZumbachLynch,LynchZumbach}:
\[
   \rho_\sigma(\dts, \dts') = \rho\left(\sigma_h[\dts, \dts/24](t), 
	\sigma_r[\dts', \dts'/24](t)\right)
\]
where on the right hand side $\rho(x, y)$ is the usual linear correlation between two time series $x$ and $y$.
Essentially, this quantity measures the dependency between past and 
future volatilities, at the respective time horizons $\dts$ and $\dts'$. 
This correlation proves to be a very powerful tool to investigate the structure of the underlying market;
in particular, it shows that the markets are heterogeneous as structures can be observed 
at the natural human time horizons (intra-day, day, week and month). 
Overall, this correlation is asymmetric with respect to the exchange of $\dts$ and $\dts'$.
As this exchange is directly related to the time reversal symmetry,
a {\em historical versus realized volatility correlation asymmetry } is defined by
\[
   a_\sigma(\dts, \dts')  = \rho_\sigma(\dts, \dts')  - \rho_\sigma(\dts', \dts).
\]
and $a_\sigma \simeq 0$ indicates time reversal invariance. 
The computation of $a_\sigma$ for the above empirical time series reveals a fairly consistent pattern, 
with a maximum of order 6 to 12\% for $\dts \simeq$ 1 week and $\dts' \simeq $ 6 hours.
The natural representation for $a_\sigma(\dts, \dts')$ is in a two dimensional plane,
but the main behaviour can be capture in a 1 dimensional cut, 
as given in fig.~\ref{fig:antisymCut_volH_volR}.
The empirical data show a distinct and fairly consistent pattern indicating 
a clear asymmetry with respect to time reversal.

\begin{figure}
  \centering
  \includegraphics[width=0.80\textwidth]{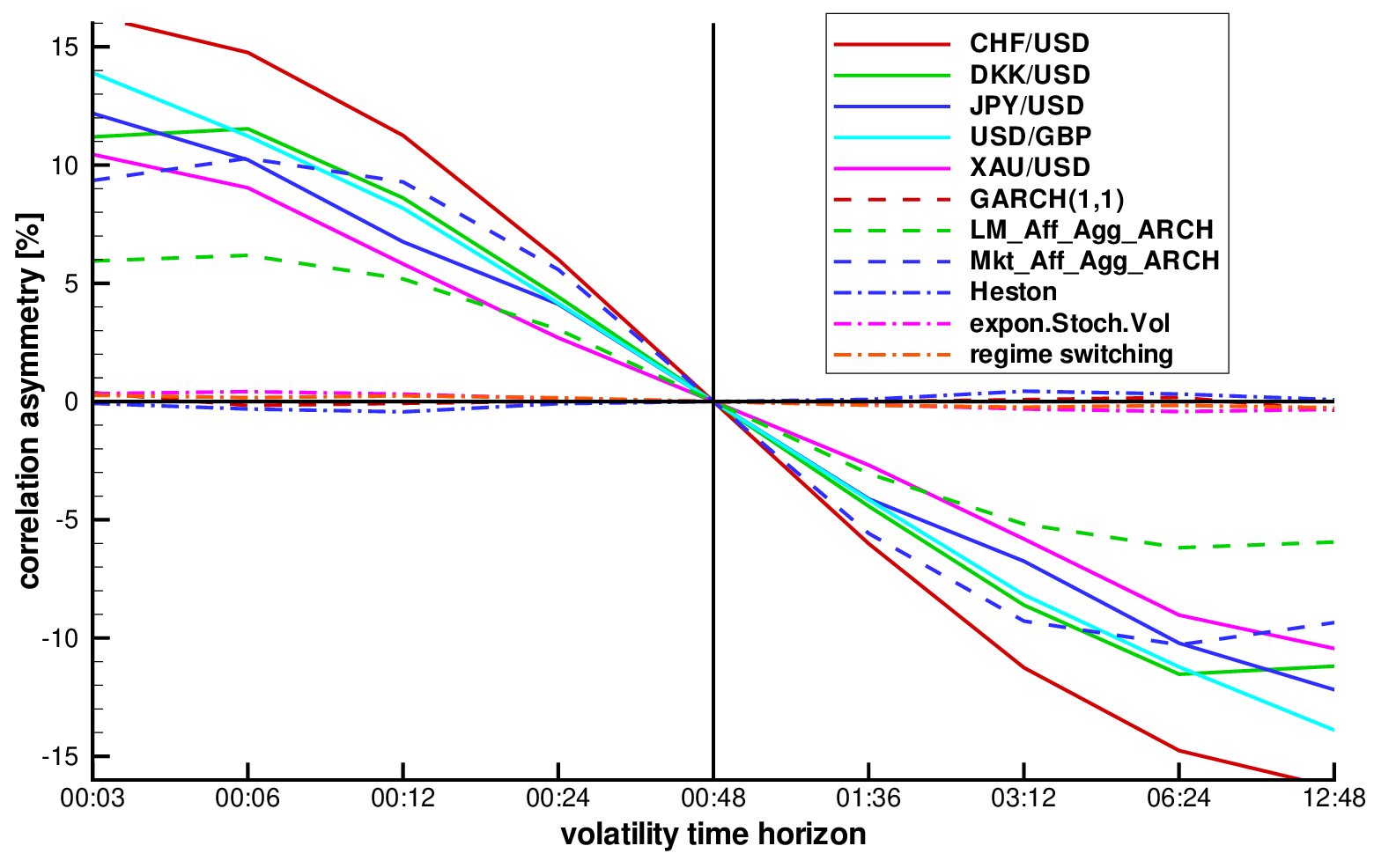}
  \caption{\it The measure of asymmetry $a_\gr(\dtr, \dtr')$.
   The parameters are: $\dts = 2^9 \cdot 3 $ minutes = 1 days 1h36; 
	$\dtr' = 2^n \cdot 3 $ minutes is given on the horizontal axis; 
   	$\dtr = 2^{8 - n} \cdot 3 $ minutes is the symmetric value through the middle point
	(for example to $\dtr'$ = 24 minutes corresponds $\dtr$ = 1h36).
   }
  \label{fig:antisymCut_volGraining}
\end{figure}
The third statistics involves correlations of past and future volatilities at the same time horizon $\dts$, 
but different granularities $\dtr$ and $\dtr'$ respectively. 
For a given $\dts$, the ``granularity'' dependency is defined by
\[
   \rho_\gr(\dtr, \dtr') = \rho(\sigma_h[\dts, \dtr](t), 
		\sigma_r[\dts, \dtr'](t))
\]
with $\dtr \leq \dts$, $\dtr' \leq \dts$. 
In \cite{Daco.1998,IntroHighFreqFin} a similar statistics was introduced, 
showing the asymmetry between fine and coarse volatilities.
As for the second statistics, the exchange of $\dtr$ and $\dtr'$ is related to the time reversal symmetry, 
and a similar measure of {\em granularity asymmetry} is defined by 
$a_\gr(\dtr, \dtr')  = \rho_\gr(\dtr, \dtr')  - \rho_\gr(\dtr', \dtr)$.
The computation of $a_\gr$ for empirical time series at $\dts \simeq$ 2 days 
shows a consistent and systematic asymmetry, 
with values in the range of 12 to 18\%.
The natural representation is in the two dimensional space $(\dtr, \dtr')$, 
and a one dimensional cut is displayed on figure~\ref{fig:antisymCut_volGraining}.
The asymmetry pattern is clear and consistent between the empirical data.
Therefore, the three measures of asymmetry deliver a consistent message, 
namely foreign exchange time series are not time reversal invariant, 
and the asymmetry is quantatively small. 

\section{TRI in theoretical processes}
A similar study can be conducted for theoretical processes, using Monte Carlo simulations. 
The simplest process is a Gaussian random walk, 
that is exactly time reversal invariant 
(the proof follows from the independence of 
the increments that allows to reorder terms under the expectations).
The more interesting processes include heteroscedasticity, 
either with an ARCH form or with a stochastic volatility term
(see e.g. \cite{Poon} for a recent general reference).

In an ARCH process, volatility is a function of previous returns.
This function can be fairly general and include for example multiple time horizons, 
or various powers of the returns ($r^2$, $|r|$, $\ln(|r|)$, ...).
An investigation among several ARCH processes shows that they are not time reversal invariant.
The simplest GARCH(1,1) process \cite{Engle.1982,Bollerslev.1986, Engle.1986} exhibits an 
asymmetry according to the measures 1 and 2, 
but the last measure displays no asymmetry 
(see figs.~\ref{fig:DV_PDF_asymmetry}, \ref{fig:antisymCut_volH_volR} 
and \ref{fig:antisymCut_volGraining}).
The multiple time horizon processes introduced in 
\cite{ZumbachLynch,LynchZumbach,Zumbach.2004} can reproduce quantitatively the above 
three measures of time irreversibility.
A key ingredient to put in these multiscale processes is that the return time horizons $\dtr$
must increase with the volatility time horizon $\dts$.
This corresponds to the intuition that short term intra-day traders use tick-by-tick data, 
whereas long term fund managers use daily data.
When $\dtr$ is kept at the process time increment, the time reversal asymmetry is too small, 
or even zero for the third measure. 
Taking a volatility granularity of order $\dts = 24\cdot\dtr$ gives roughly the 
correct quantitative time reversal asymmetry in these multiple components processes.
On the other extreme, taking $\dts = \dtr$ produces too much asymmetry.
Clearly, the process parameters can be chosen in order to reach a 
better numerical agreement with the empirical values (no such optimisation has been done for this work).

In a stochastic volatility (SV) process, 
the volatility is an independent process with its own source of randomness
(see e.g. \cite{Shephard} for a general reference).
The return is a ``slave'' process dependent on the volatility, and there is
no feed-back from the return to the volatility.
The volatility process can be fairly simple, 
as in an exponential SV process or as in the Heston process.
In the exponential SV process, 
the logarithm of the volatility follows an Ornstein-Uhlenbeck process.
In the Heston process, the volatility follow a random walk with mean revertion, 
but the stochastic term is multiplied by $\sqrt{\sigma}$ so that $\sigma$ stays positive.
Both processes can also be extended to include multiple time horizons 
in order to induce a volatility cascade over multiple time scales.
Two such models with a long memory volatility cascade 
are label in the graphs and tables as ``LM stoch.vol.'' and ``LM Heston''.
Yet, all these SV processes show time reversal invariance 
(up to the statistical errors of the numerical simulation, 
see figs.~\ref{fig:DV_PDF_asymmetry}, \ref{fig:antisymCut_volH_volR} 
and \ref{fig:antisymCut_volGraining}).

\begin{figure}
  \centering
  \psfrag{X}{$\Delta\sigma$}
  \psfrag{Y}{$a_p(\Delta\sigma)$}
  \includegraphics[width=\figWidth]{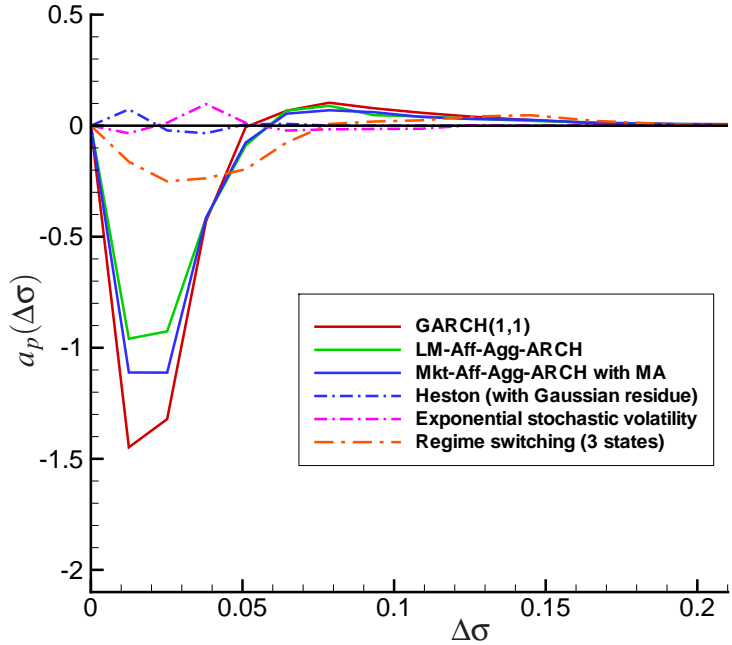}
  \caption{\it Asymmetry of the probability density for the volatility 
  increment $\Delta\sigma$ for a few processes.
  }
  \label{fig:process_PDF_asymmetry}
\end{figure}

A family of models with a similar structure is the regime switching processes. 
In this case, the volatility is given by another independent process with an integer value 
that gives the ``state of the world'' (like quiet, excited, shocked, etc...).
To each index value corresponds a volatility, 
and the return follows a simple random walk with the corresponding volatility.
The state of the world is an independent process, without feed-back from the return, 
and follows a simple Markov process with probability $\Prob(i\rightarrow j)$
to jump from state $i$ to state $j$.
The transition probabilities $\Prob(i\rightarrow j)$ are strongly constrained
to give realistic distribution for the volatility.
With asymmetric transition probabilities $\Prob(i\rightarrow j) \neq \Prob(j\rightarrow i)$,
the process is not time reversal invariant according to the measures $a_p$ and $a_\sigma$,
as is visible on fig.~\ref{fig:process_PDF_asymmetry}.
Yet, the asymmetry is much smaller than the empirical observed values, and it is not possible to modify the transition probabilities to get simultaneously 
large asymmetry and realistic pdf.
Finally, the coarse/fine graining measure of asymmetry $a_\gr$ for the regime switching
process is compatible with zero, up to Monte Carlo statistical fluctuations. 
Because of the lack of feed-back of the return on the volatility, one can understand intuitively that the stochastic volatility and regime switching cannot include an asymmetry in the third measure of invariance $a_\gr$ using the volatility granularity.

\section{Test statististics}
So far, the focus was to construct statistics sensitive to time reversal invariance,
to understand what they are measuring,
and to show the differences between empirical time series and various processes.
Yet, definitive conclusions require to use rigorous test statistics.
Essentially three methods can be followed in order 
to obtain distribution information on a statistics: 
analytical, bootstrap and Monte Carlo simulations.
In order to select the most appropriate method for the asymmetry statistics, 
one should keep in mind that our statistics are based on volatilities 
and that volatilities have slowly decaying correlations (like a power law).
The bootstrap method is appropriate only when the data are independent, 
or possibly when the dependencies decay exponentially fast.
Clearly, the slow decay of the correlations rules out a bootstrap approach.
Similarly, the analytical approaches are using a convergence toward a limit law,
like a Gaussian. 
The convergence needs to be sufficiently fast, 
and similar conditions are imposed on the data as for the bootstrap method.
Moreover, the asymmetry statistics are not based on simple moments, 
but on probability distributions and correlations. 
Both problems makes the analytical approach fairly difficult.
Therefore, we have used Monte Carlo simulations to compute numerically the distributions 
for the asymmetry statistics and the related $p$-values.
The computed $p$-values are the probability that the statistics is lower or equal to zero.
For a symmetic distribution, we expect a $p$-value around 0.5.

In order to obtain the distribution of the asymmetry statistics for a given process,
200 Monte Carlo simulations are performed.
For each simulation, 
the process is simulated with a time increment of 3 minutes and 
for a time length identical to the available empirical data, namely 11.5 years.
At the end of the simulation, asymmetry statistics are computed.
This is repeated 200 times, and the empirical distributions for the asymmetry statistics are computed.
The cumulative probability that a statistic is smaller or equal to zero can be easily estimated, 
and gives the desired $p$ value.
This approach is simple and gives finite sample information, 
but relies on the process that should reproduce sufficiently well the empirical data.

The distribution statistics can be obtained for given values of the arguments, 
say for example $a_p(\Delta\sigma)$ for a given $\Delta\sigma$.
Yet, the sensitivity of the TRI test can be enhanced by simple integration 
of the above statistics (similarly to a portmanteau statistics for the correlation).
The {\em volatility increment asymmetry} is defined by
\begin{equation}
	A_p(\Delta\sigma) = \frac{1}{n} \sum_{0 < \Delta\sigma_k < \Delta\sigma} a_p(\Delta\sigma_k) 
\end{equation}
where $\Delta\sigma_k$ are the points on the histogram sampling grid
and where $n$ is the number of terms in the sum.
In the computations below, the upper bound is $\Delta\sigma$ = 0.06.
For the historical-realized volatility correlation asymmetry, two possibilities are
\begin{eqnarray}
	A_{\sigma, \text{tot}} & = & \int_{\dt_{\min}}^{\dt_{\max}} d\dts ~\int_{\dt_{\min}}^{\dt_{\max}} d\dts' ~~a_\sigma(\dts, \dts') \\
	A_{\sigma, \text{cut}} & = & \int_{\dt_{\min}}^{\dt_{\max}}  d\dts ~a_\sigma(\dts, f(\dts)).
\end{eqnarray}
The first quantity computes the total asymmetry in the historical-realized volatility,
the second one the asymmetry along the cut used for figure~\ref{fig:antisymCut_volH_volR}.
In practice, the two statistics give very similar results.
Table~\ref{table:A_sigma_cut} reports the $p$-values for $A_{\sigma, \text{cut}}$ 
as the values are related to figure~\ref{fig:antisymCut_volH_volR}.
For the volatility graining asymmetry, the same two integrated measures can be constructed.
Again to stay close to figure~\ref{fig:antisymCut_volGraining}, the statistics for 
\begin{equation}
	A_{\gr, \text{cut}} = \int_{\dt_{\min}}^{\dt_{\max}} \dtr ~~a_\gr(\dtr, f(\dtr))
\end{equation}
are reported.

\begin{table}
\centering
\begin{tabular}{|r|c|c|c|}
\hline
                  & mean & stdDev & p-value \\
\hline
CHF/USD           & 0.64   &        &        \\
DKK/USD           & 0.39   &        &        \\
JPY/USD           & 0.65   &        &        \\
USD/GBP           & 0.95   &        &        \\
XAU/USD           & 0.63   &        &        \\
\hline
GARCH(1,1)        & 0.19   & 0.083  & 0.014  \\ 
LM-Aff-Agg-ARCH   & 0.26   & 0.082  & 0.015  \\
Mkt-Aff-Agg-ARCH  & 0.39   & 0.1    & 0.0    \\ 
\hline
exp stoch.vol.    & 0.062  & 0.12   & 0.35   \\
exp LM stoch.vol. & -0.02  & 0.09   & 0.48   \\ 
Heston            & 0.008  & 0.075  & 0.36   \\
LM Heston         & 0.02   & 0.1    & 0.50   \\ 
Regime Switching  & 0.23   & 0.11   & 0.015  \\
\hline
\end{tabular}
\caption{
The total measure $A_p$ for the asymmetry of the probability density of $\Delta\sigma$
at a time horizon of $\dts = $ 1 day.
The columns give respectively the mean, standard deviation and $p$-value for $A_p$.
\label{table:A_p}
}
\end{table}


\begin{table}
\centering
\begin{tabular}{|r|c|c|c|}
\hline
                  & mean   & stdDev & p-value \\
\hline
CHF/USD           & 0.1    &        &        \\
DKK/USD           & 0.09   &        &        \\
JPY/USD           & 0.05   &        &        \\
USD/GBP           & 0.09   &        &        \\
XAU/USD           & 0.08   &        &        \\
\hline
GARCH(1,1)        & 0.007  & 0.02   & 0.42   \\ 
LM-Aff-Agg-ARCH   & 0.14   & 0.01   & 0.0    \\ 
Mkt-Aff-Agg-ARCH  & 0.15   & 0.01   & 0.0    \\ 
\hline
exp stoch.vol.    & 0.0009 & 0.01   & 0.50   \\ 
exp LM stoch.vol. &-0.0016 & 0.01   & 0.52   \\ 
Heston            & 0.0005 & 0.007  & 0.49   \\ 
LM Heston         &-0.001  & 0.01   & 0.54   \\ 
Regime Switching  & 0.004  & 0.01   & 0.34   \\ 
\hline
\end{tabular}
\caption{
The integrated measure of asymmetry for the historical/realized volatility correlations
$A_{\sigma, \text{cut}}$ along a cut in the $(\dts, \dts')$ plane.
\label{table:A_sigma_cut}
}
\end{table}


\begin{table}
\centering
\begin{tabular}{|r|c|c|c|}
\hline
                  & mean    & stdDev & p-value \\
\hline
CHF/USD           & 0.13    &        &        \\
DKK/USD           & 0.1     &        &        \\
JPY/USD           & 0.095   &        &        \\
USD/GBP           & 0.1     &        &        \\
XAU/USD           & 0.08    &        &        \\
\hline
GARCH(1,1)        & 0.0008  & 0.007  & 0.48   \\ 
LM-Aff-Agg-ARCH   & 0.1     & 0.01   & 0.0    \\ 
Mkt-Aff-Agg-ARCH  & 0.11    & 0.01   & 0.0    \\ 
\hline
exp stoch.vol.    &-0.00007 & 0.008  & 0.58   \\ 
exp LM stoch.vol. &-0.0007  & 0.006  & 0.54   \\ 
Heston            & 0.00007 & 0.01   & 0.52   \\ 
LM Heston         & 0.0003 & 0.007  & 0.59   \\ 
Regime Switching  & 0.0002  & 0.01   & 0.46   \\ 
\hline
\end{tabular}
\caption{
The integrated measure of asymmetry for the volatility graining correlation
$A_{\gr, \text{cut}}$ along a cut in the $(\dtr, \dtr')$ plane.
\label{table:volGrainingAsymmetry_cut}
}
\end{table}

\begin{figure}
  \centering
  \includegraphics[width=0.80\textwidth]{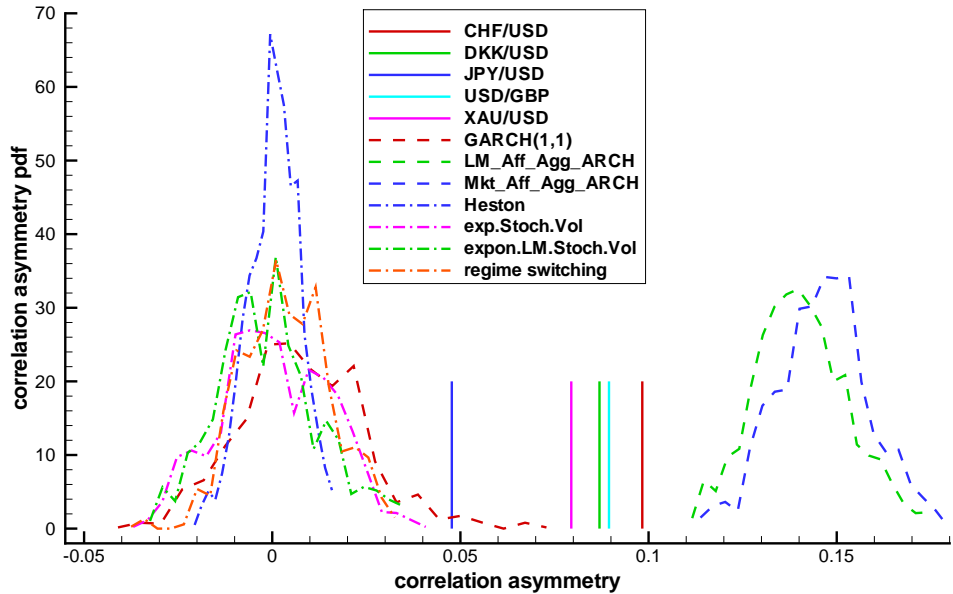}
  \caption{\it The probability density for the asymmetry measure $A_{\sigma, \text{cut}}$.
  }
  \label{fig:histRealVolCorr_cut_PDF_asymmetry}
\end{figure}
The main statistics are reported in the table \ref{table:A_p} to \ref{table:volGrainingAsymmetry_cut}.
For the historical versus realized volatility correlation asymmetry, 
the probability distributions for the processes are given in figure~\ref{fig:histRealVolCorr_cut_PDF_asymmetry}.
The $p$-statistics confirm fully the above picture:
the empirical data are compatible only with multicomponent ARCH processes. 
Notice that the process parameters have not been optimized to reproduce quantitatively the empirical asymmetry.
A better agreement can be obtained for the multicomponent 
ARCH processes as they contain parameters that control the asymmetry.
This is not possible for other stochastic volatility processes, 
say for example with respect to the volatility graining asymmetry: 
regardless of the parameter values, the $p$-values stay close to 1/2.
Notice also that a simple GARCH(1,1) process can reproduce the asymmetry for $a_p$, 
but the asymmetry for $a_\sigma$ is too small and the asymmetry for $a_\gr$ is not reproduced at all.
 
\begin{table}
\centering
\begin{tabular}{|r|c|c|c|}
\hline
                  &  mean   & stdDev & p-value \\
\hline
CHF/USD           & -0.21   &       &       \\
DKK/USD           & -0.23   &       &       \\
JPY/USD           & -0.11   &       &       \\
USD/GBP           & -0.08   &       &       \\
XAU/USD           &  0.21   &       &       \\
\hline
GARCH(1,1)        & -0.009  & 0.07   & 0.54  \\ 
LM-Aff-Agg-ARCH   & -0.005  & 0.07   & 0.45  \\
Mkt-Aff-Agg-ARCH  & -0.0009 & 0.09   & 0.44  \\ 
\hline
exp stoch.vol.    & -0.002  & 0.08   & 0.42  \\
exp LM stoch.vol. &  0.0015 & 0.08   & 0.54  \\ 
Heston            & -0.002  & 0.075  & 0.50  \\
LM Heston         &  0.002  & 0.08   & 0.55  \\ 
Regime Switching  & -0.005  & 0.08   & 0.59  \\
\hline
\end{tabular}
\caption{
The total measure of the asymmetry of the probability 
density of $r$ at a time horizon of 1 day.
The columns give respectively the mean, standard deviation and $p$-value.
\label{table:return}
}
\end{table}

\section{Possible origins of time irreversibility}
With the help of both the empirical analysis and the process structures, 
we can speculate on the origin of the time asymmetry.
The systematic study on the processes indicates that the time direction is set by the feed-back loop
of the price changes on themselves through the intermediary of the price volatility, 
as captured by an ARCH process. 
Moreover, to obtain the measures of asymmetry with the right magnitudes, 
multiple time scales must be used in an ARCH structure.
This indicates that the interplay between the different time horizons as well as the feed back loop 
through the price are enough to create the observed time asymmetry.
This argument emphasises the role of the prices as the only vector of information between market participants.
By contrast, other ``hidden'' variables 
such as an independent stochastic volatility or other market states are irrelevant with respect to TRI.
The picture that emerges is of markets segmented along the time horizons of the market participants.
A more detailed analysis of the historical versus realized volatilities points in the same direction,
with a cascade from short time horizons to long horizons \cite{ZumbachLynch,LynchZumbach}.

Another relevant topic with respect to TRI is the connection between news and market behaviour.
Clearly, financial markets are open systems, driven by various external sources of information,
like exchange of goods, politics, central banks, etc...
Even if intuitively clear, the connection between ``news'' and financial markets 
is very difficult to establish from a purely empirical point.
The two fundamental problems are the economic quantification of a ``string of text'', 
and the discounting by market participants of the ``expected'' portion of a news.
With respect to TRI, the important question is whether the irreversibly is
of external origin (i.e. from the news) or endogenous 
(i.e. created by the market participants or the trading rules).
As such, this question is very difficult to investigate empirically,
but the present study offers some indirect evidence.
In the equations for a process, there is at least one source of randomness,
for example in the return equation $r(t) = \sigma(t)~\epsilon(t)$.
Possibly, a process can include several sources of randomness, 
like in a stochastic volatility process.
Essentially, the random variables $\epsilon(t)$ 
capture two different phenomenon: 
the trades of the market participants (for example through an order queue),
and the influence of the external world (i.e. the news).
The canonical hypothesis for the random processes $\epsilon$ is to assume 
an iid distribution (and independence at a given time 
with the other process variables
like price, return, volatility, etc ...).
A debatable assumption is that of time series independence: 
it is plausible that the news are serially correlated, 
as an important piece of information is likely 
to be followed by more information.
But this is difficult to establish directly, 
again because news are given by 
texts and not by numbers.
In the investigation of processes, 
we follow the canonical hypothesis of source of randomness that are iid.
Within this hypothesis, two processes are particularly interesting with respect to TRI.
In a regime switching process, the ``state of the world'' $i(t)$
exhibits irreversibility, as controlled by the transition 
probabilities $\Prob(i\rightarrow j)$, and $i(t)$ has non trivial serial correlations.
For this process, a parallel can be established with the serial correlation of the news.
Yet, it is not possible with a regime switching process to reproduce the observed time asymmetry.
On the other hand, within the framework of iid source of randomness, 
the multiscale GARCH processes can provide
a quantitatively correct description of the time irreversibility observed in the 
empirical time series.
Together, both arguments indicate that the time irreversibility is mostly of endogenous origin, 
namely created by the interactions between market participants,
and not from external origin.

In a physical system, a common origin of time irreversibility is in friction or dissipation, 
and in the increase of entropy.
A direct analogy of these quantities in a financial system should be taken with care.
For example, the equivalent of friction can be thought as the bid-ask spread, 
namely the cost for buying, followed by selling directly afterward.
However, in a liquid markets, and particularly in modern electronic trading exchanges, 
the market orders and limit orders play a very symmetric role.  
Using an argument based on the optimal choice of the market agents between both kind of orders, 
the authors in \cite{Bouchaud.2006} show that the profit of a 
systematic market making trading strategy should be close to zero.
Indirectly, this shows that the raw cost of trading represents a very small dissipation in the system.
In the same direction, all the simulated processes do not include the cost of transactions,
yet all the ARCH processes are not TRI.
Both arguments show that the observed asymmetry in time does not originate in the bid-ask spread.

Notice that the above discussion on the origin of time irreversibility presents only indirect arguments, 
as only an investigation of the market 
microstructure and of the market participant decision processes can give direct evidence.
Yet, such studies are clearly very difficult, and even more on 
large decentralised markets as foreign exchanges.

All the measures of asymmetry used so far are even in the returns, 
and essentially related to volatilities.
The asymmetry in the return distribution is another measure of time asymmetry, 
which is odd in the returns.
As discussed in the introduction, 
this symmetry is also related to the exchange of currencies for the foreign exchange time series.
The empirical results for the return distribution asymmetry are reported in table~\ref{table:return}.
All the processes have rigorously symmetric return distributions 
(because the returns are proportional to the residuals, which have a symmetric distributions).
The results for the processes are in clear agreement with this symmetry, with p-values close to 1/2.
Interestingly, the asymmetry for the empirical data is much larger
than for any processes (roughly at 1 $\sigma$ to 2.5 $\sigma$ from 0).
This indicates that the argument on the exchange of currencies in a FX rate, 
as given in the introduction, is likely not correct.
Clearly, more extensive investigations are needed to clarify the empirical facts 
and to develop processes that can reproduce the observed return asymmetry in the data.

\section{Conclusions}
To conclude, the systematic study of the time reversal invariance 
in finance proves to be a very powerful investigation tool.
The empirical time series are clearly not time reversal invariant, according to three different measures.
Although this fact is not obvious at the level of the price time series, 
this is not completely surprising as markets are driven by humans, 
who are clearly not time reversal invariant.
In particular, the market participants remember the past, 
and this memory creates an asymmetry in the time direction.
More interesting is the fact that different FX rates show a clear and consistent quantitative pattern.
On the modelling side, the ARCH processes can accommodate the empirical finding, 
using multiscale processes with increasing return time horizons $\dtr$ 
for increasing volatility time horizons $\dts$.
The simplest GARCH(1,1) process can only reproduce some asymmetry, 
because it contains only one time scale.
On the other hand, all the stochastic volatility and regime switching processes 
are essentially time reversal invariant, at odds with the empirical data. 
This deficiency is related to the structure of these processes, 
and is therefore a key shortcoming of this class of models.
Eventhough the various theoretical processes have very different structures, 
and in particular are not nested, 
TRI provides us with a strong selection criterion: 
a large number of processes cannot reproduce the stylized facts 
related to the time irreversibility observed in empirical time series. 

\section{Acknowledgements}
The author thanks Jean-Philippe Bouchaud and Pascal deRougemont for their questions and discussions
on time reversal in finance that initiated this paper.

\newpage
\bibliographystyle{apalike}

\end{document}